\documentclass{ifacconf}
\usepackage{graphicx,graphics}
\usepackage[config, labelfont={sf,bf}, textfont=sf]{caption,subfig}
\graphicspath{{./figs/}}
\usepackage{natbib}        
\usepackage{amsmath,amsfonts}   

\newtheorem{theorem}{Theorem}

\usepackage[suffix=, outdir=./]{epstopdf}

\begin{document}
\begin{frontmatter}

\title{Localization of hidden Chua attractors by the describing function method}

\author[spbu,jyv]{N.V. Kuznetsov},
\author[spbu]{O.A. Kuznetsova},
\author[spbu,ipme]{G.A. Leonov},
\author[spbu]{T.N. Mokaev},
\author[jyv,sstu]{N.V. Stankevich}

\address[spbu]{Faculty of Mathematics and Mechanics, St. Petersburg State University, Russia}
\address[ipme]{Institute of Problems of Mechanical Engineering RAS, Russia}
\address[jyv]{Dept. of Mathematical Information Technology,
University of Jyv\"{a}skyl\"{a}, Finland}
\address[sstu]{Faculty of Electronics and Instrumentation,
Saratov State Technical University, Russia}

\begin{abstract}
    In this paper the Chua circuit with five linear elements and saturation non-linearity is studied.
    Numerical localization of self-excited attractor in the Chua circuit model
    can be done by computation of trajectory with initial data in a vicinity of an
    unstable equilibrium.
    For a hidden attractor its basin of attraction
    does not overlap with a small vicinity of equilibria, so it is difficult to find
    the corresponding initial data for localization.
    This survey is devoted to the application of describing function method
    for localization of hidden periodic and chaotic attractors in the Chua model.
    We use a rigorous justification of the describing function method,
    based on the method of small parameter,
    to get the initial data for the visualization of the hidden attractors.
    A new configuration of hidden Chua attractors is presented.
\end{abstract}


\end{frontmatter}

\section{Introduction}

In the initial period of the development of
the theory of nonlinear oscillations (first half of the XX century)
main attention of researchers was paid to analysis
and synthesis of oscillating systems
for which the oscillation existence problem can be solved
relatively easily.
The structure of many applied systems considered was such that the existence
of oscillations was "almost obvious" - the oscillation was excited
from an unstable equilibrium (so called {\it self-excited oscillation}).
From a computational point of view this allows one to use a {\it standard computational procedure},
in which after a transient process a trajectory, started from a point of unstable
manifold in a neighborhood of equilibrium, reaches a state of oscillation,
therefore one can easily identify it.
The use of the term \emph{self-excited oscillation} or {\it self-oscillations}
can be traced back to the works of H.G.~Bark\-hausen and A.A.~Andronov,
where it describes the generation and maintenance of a periodic motion in mechanical
and electrical models by a source of power that lacks any corresponding periodicity
(e.g., a stable limit cycle in the van der Pol oscillator) \citep{AndronovVKh-1966,Jenkins-2013}.

Attractor is called a \emph{self-excited attractor} if its basin of attraction
intersects any arbitrarily small open neighborhood of an equilibrium,
otherwise it is called a \emph{hidden attractor}
\citep{LeonovKV-2011-PLA,LeonovKV-2012-PhysD,LeonovK-2013-IJBC,LeonovKM-2015-EPJST,Kuznetsov-2016}.
We use the notion ``self-excited'' for attractors of dynamical systems
to describe the existence of transient process
from a small vicinity of an unstable equilibrium
to an attractor.

If  there is no such a transient process for an attractor,
it is called a hidden attractor.
For example, hidden attractors are attractors in systems
without equilibria or with only one stable equilibrium
(a special case of multistability and coexistence of attractors).
Some examples of hidden attractors can be found in
\cite{ShahzadPAJH-2015-HA,BrezetskyiDK-2015-HA,JafariSN-2015-HA,ZhusubaliyevMCM-2015-HA,SahaSRC-2015-HA,Semenov20151553,FengW-2015-HA,Li20151493,FengPW-2015-HA,Sprott20151409,PhamVVJ-2015-HA,VaidyanathanPV-2015-HA,
Danca-2016-HA,Zelinka-2016-HA,DudkowskiJKKLP-2016,KuznetsovLYY-2017-CNSNS,DancaKC-2016,KiselevaKL-2016-IFAC}.

\begin{figure}[!ht]
  	\centering
    	\includegraphics[width=0.4\textwidth]{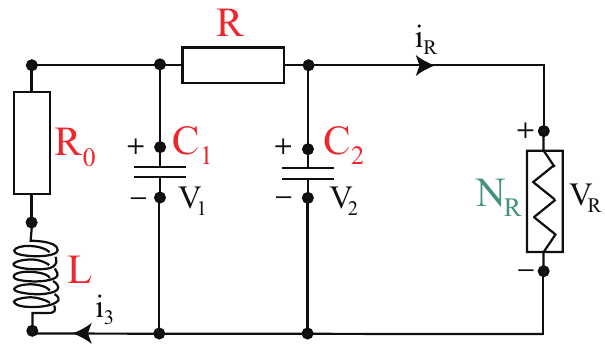}
    \caption{Chua circuit with two resistors, one inductor, two capacitors (red)
    and one nonlinear resistor called ``Chua diode'' (green).}
    \label{fig:chuacircuit}
\end{figure}

\begin{figure}[!ht]
  \centering
  \subfloat[Parameters $\alpha = 15$, $\beta = 28$, $\gamma = 0$, $m_0 = -5/7$, $ m_1 = -8/7$.]{
    \label{fig:self-excited:spiral}
    \includegraphics[width=0.23\textwidth]{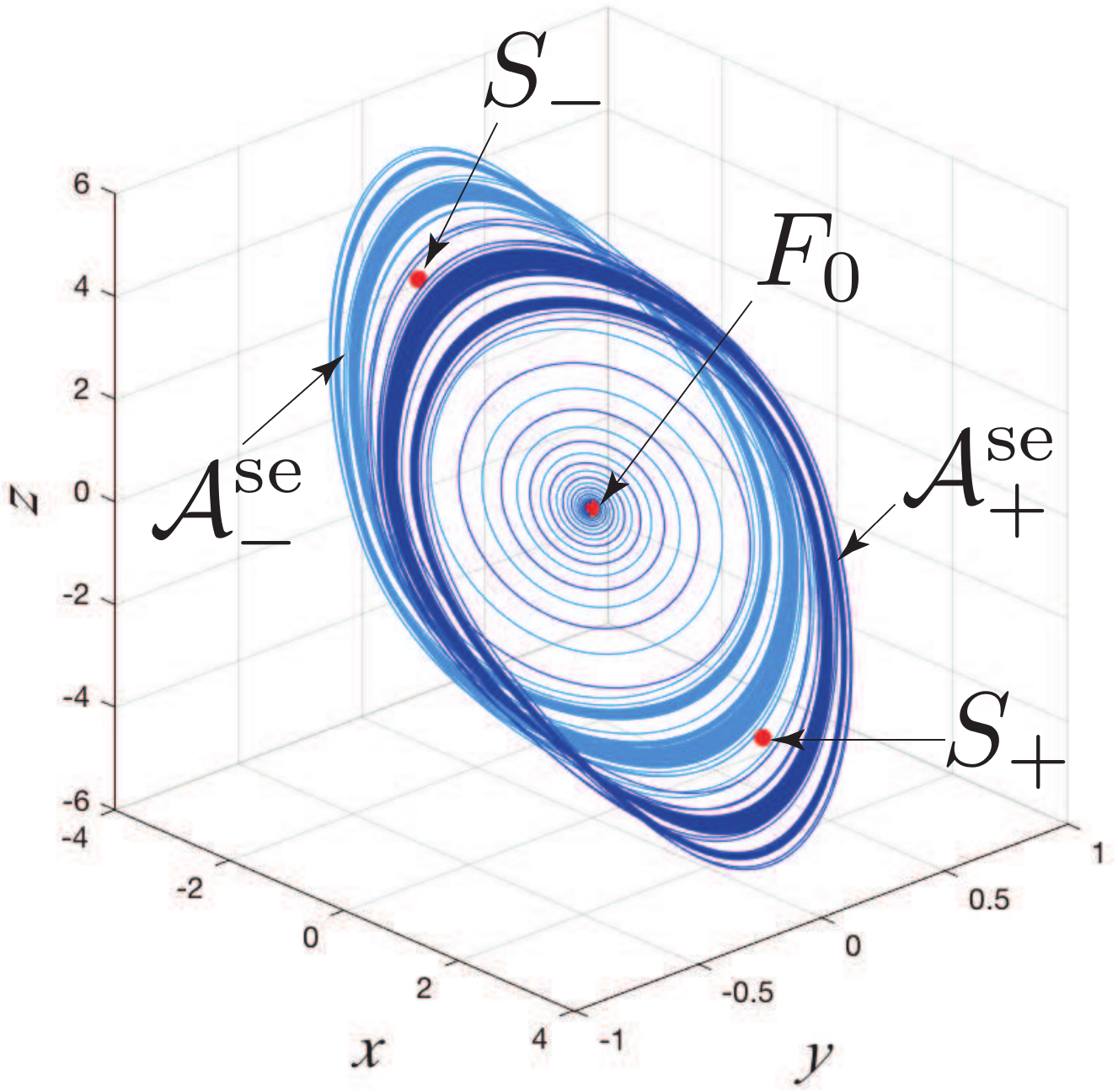}
  }~
  \subfloat[Parameters $\alpha = 8.5$, $\beta = 14.28$, $\gamma = 0$, $m_0 = -8/7$, $ m_1 = -5/7$.]{
    \label{fig:self-excited:rt}
    \includegraphics[width=0.23\textwidth]{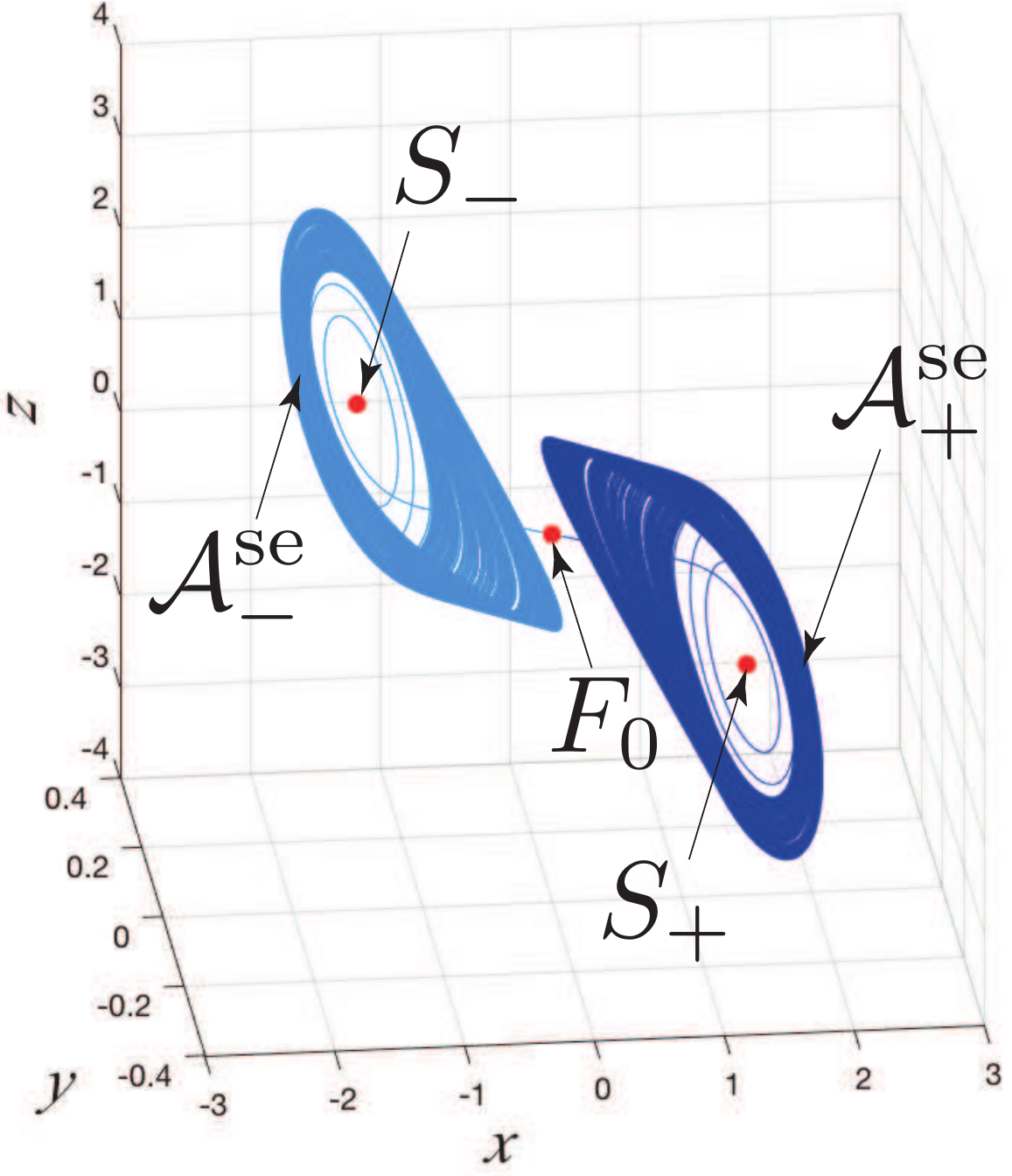}
  }
  \caption{Self-excited attractors in Chua system \eqref{chuasys}:
  \eqref{fig:self-excited:spiral} -- two symmetric spiral attractors,
  \eqref{fig:self-excited:rt} -- two symmetric R\"{o}ssler-like attractors.
  }
  \label{fig:self-excited}
\end{figure}

The \emph{self-excited and hidden classification of attractors}
was introduced by Leonov and Kuznetsov in connection with the discovery of hidden chaotic attractor
in the Chua system \citep{KuznetsovLV-2010-IFAC,LeonovKV-2011-PLA,KuznetsovKLV-2013}:
\begin{equation}\label{chuasys}
     \begin{aligned}
         \dot x&=\alpha(y-x(m_1+1))-\alpha\psi(x),\\
         \dot y&=x-y+z,\\
         \dot z&=-(\beta y+\gamma z),\\
         \psi(x)&= (m_0-m_1) \, {\rm sat}(x) = \\
         &=\frac{1}{2}(m_0-m_1)(|x+1|-|x-1|),
     \end{aligned}
 \end{equation}
where $\alpha$, $\beta$, $\gamma$, $m_0$, $m_1$ are parameters.
This system provides a mathematical model, describing the behavior of the Chua circuit
\citep{Chua-1990,Chua-1992,Chua-1995} with five linear elements and saturation non-linearity
(see Fig. \ref{fig:chuacircuit}).
Until this discovery only self-excited chaotic attractors had been found in Chua circuits
(see Fig.~\ref{fig:self-excited} and, e.g. works \citep{Matsumoto-1984,Lozi-1993,BilottaP-2008}).
Note that L. Chua himself, analyzing various cases of attractors existence in Chua circuit,
does not admit the existence of hidden attractor in his circuits \citep{Chua-1992}.

We consider only one type of the Chua circuits,
while there are known various modifications of Chua circuit
(see, e.g. \citep{Banerjee-2012,Semenov20151553})
where hidden oscillations can also be localized
\citep{ChenYB-2015-HA,BaoHCXY-2015-HA,ChenLYBXW-2015-HA,MenacerLC-2016-HA}.

\section{Hidden attractors localization via describing function method}

In this section an effective analytical-numerical approach
for hidden oscillations localization, based on
the describing function method (DFM),
the method of small parameter
and continuation method, is demonstrated.

\subsection{Describing function method}

The describing function method (DFM)
is a searching method for oscillations which are close
to the harmonic periodic oscillations of nonlinear systems of automatic control.
This method is {\it not strictly mathematically justified}
and is one of approximate methods of analysis of control systems
(see, e.g. \citep{KrylovB-1947,Khalil-2002}).
One of the first examples, where the describing function method
gives untrue results, is due to \cite{Tzipkin-1984}.
Remark that well-known Aizerman's and Kalman's conjectures
on the absolute stability of nonlinear control systems
are valid from the standpoint of
the describing function method
(what explains why these conjectures were put forward).
Nowadays various counterexamples to these conjectures
(nonlinear systems where the only equilibrium, which is stable,
coexists with a hidden periodic oscillation) are known
(see, e.g. \cite{Pliss-1958,Fitts-1966,Barabanov-1988,BernatL-1996,LeonovBK-2010-DAN,BraginVKL-2011,LeonovK-2011-DAN,LeonovK-2013-IJBC};
the corresponding discrete examples are considered in \cite{Alli-Oke-2012-cu,HeathCS-2015}).

Let us recall a classical way of applying the DFM.
Consider a system with one scalar non-linearity in the Lur'e form
 \begin{equation}\label{sys_gen}
   \frac{d {\bf x}}{dt}={\bf P}{\bf x}+{\bf q}\psi({\bf r}^*{\bf x}),\quad
   {\bf x}\in \mathbb{R}^n
 \end{equation}
where ${\bf P}$ is a constant $(n\times n)$-matrix,
${\bf q}, {\bf r}$ are constant $n$-dimensional vectors,
$^*$ denotes transpose operation,
$\psi(\sigma)$ is a continuous piecewise-differentiable
scalar function, and $\psi(0)=0$.

In order to find a periodic oscillation,
a certain coefficient
of harmonic linearization $k$ (assume that such $k$ exists)
is introduced in such a way that the matrix
${\bf P}_0 = {\bf P} + k{\bf qr}^*$ of the linear system
\[
	\frac{d {\bf x}}{dt}={\bf P}_0{\bf x}, \quad
   	{\bf x}\in \mathbb{R}^n
\]
has a pair of pure-imaginary eigenvalues $\pm i\omega_0 (\omega_0>0)$,
and the rest eigenvalues have negative real parts.

Introduce a transfer function
\begin{equation}\label{eq:transfer_func}
	W(p) = {\bf r}^* \big({\bf P} - p {\bf I} \big)^{-1} {\bf q},
\end{equation}
where $p$ is a complex variable, ${\bf I}$ is a unit matrix.
Transfer function $W(p)$ is applied to define the values of $k$ and $\omega_0$.
The number $\omega_0 > 0$ is defined from the equation
\begin{equation}\label{eq:omega0}
	{\rm Im}\, W( i \, \omega_0) = 0
\end{equation}
and $k$ is defined by the formula
\begin{equation}\label{eq:harm-lin-coef}
	k = - \big({\rm Re}\, W(i \, \omega_0) \big)^{-1}.
\end{equation}
If such $\omega_0$ and $k$ exist,
then system \eqref{sys_gen}
has a periodic solution ${\bf x}(t)$ for which
\[
	\sigma(t) = {\bf r}^*{\bf x}(t) \approx a \cos \omega_0 t.
\]
Following the DFM, the
amplitude $a$ can be obtained from the equation
\[
	\int_{0}^{2\pi/\omega_0} \big(\psi(a \cos \omega_0 t) a \cos \omega_0 t
		- k (a \cos \omega_0 t)^2 \big) dt = 0.
\]
Rewrite system \eqref{sys_gen} as follows
\begin{equation} \label{sys_gen_p0}
  \frac{d{\bf x}}{dt}={\bf P_0}{\bf x}+{\bf q}\varphi({\bf r}^*{\bf x}),
\end{equation}
where $\varphi(\sigma)=\psi(\sigma)-k\sigma$.
As it is mentioned above classical DFM is not strictly mathematically justified
and can lead to untrue results, however
for the systems with a small parameter it can be rigorously justified.
For that let us change $\varphi(\sigma)$ by $\varepsilon\varphi(\sigma)$
and consider the existence of a periodic solution for system
\begin{equation} \label{sys_gen_phi_0}
   \frac{d{\bf x}}{dt}={\bf P_0}{\bf x}+\varepsilon{\bf q}\varphi^0({\bf r}^*{\bf x}).
\end{equation}
To define the initial data ${\bf x}^0(0)$ of the periodic solution,
system \eqref{sys_gen_phi_0} is transformed by a linear non-singular transformation ${\bf x}={\bf S}{\bf y}$
to the form\footnote{Such transformation exists for non-degenerate transfer functions.}
\begin{equation} \label{sys_hlin}
 \begin{aligned}
 & \dot y_1 =-\omega_0y_2+\varepsilon b_1\varphi(y_1+{\bf c}_3^{*} {\bf y}_3),
  \\
 & \dot y_2 = \omega_0y_1+\varepsilon b_2\varphi(y_1+{\bf c}_3^{*} {\bf y}_3),
\\
 & \dot {\bf y}_3 = {\bf A}_3{\bf y}_3+\varepsilon {\bf b}_3\varphi(y_1+{\bf c}_3^{*} {\bf y}_3),
 \end{aligned}
\end{equation}
where $y_1$, $y_2$ are scalars, ${\bf y}_3$, ${\bf b}_3$ and ${\bf c}_3$ are $(n-2)$-dimensional vectors,
$b_1$ and $b_2$ are real numbers;
${\bf A}_3$ is a constant $((n-2)\times(n-2))$ matrix
all eigenvalues of which have negative real parts.
Without loss of generality, it can be assumed that for the matrix
${\bf A}_3$ there exists a positive number $d>0$, such that
${\bf y}_3^*({\bf A}_3+{\bf A}_3^*){\bf y}_3\le-2d|{\bf y}_3|^2$,
$\forall\,{\bf y}_3\in\mathbb{R}^{n-2}$.

Introduce the describing function
\begin{equation}\label{eq:descr-func}
  \Phi(a)=\int\limits_0^{2\pi\!/\omega_0}\varphi(\cos(\omega_0t)a)\cos(\omega_0t)dt
\end{equation}
and assume the existence of its derivative.
\begin{theorem}\label{th_stable} [\cite{LeonovK-2013-IJBC}]
If there exists a positive number $a_0$ such that
\begin{equation}\label{det_a0}
  \Phi(a_0)=0, \quad b_1\frac{d\Phi(a)}{da}\bigg|_{a=a_0} < 0,
\end{equation}
then system \eqref{sys_gen_phi_0} has a stable periodic solution with initial data
\[
	{\bf x}^0(0) = {\bf S}\,\big(y_1(0), \, y_2(0), \, y_3(0)\big)^*,
\]
where $y_1(0) = a_0 + O(\varepsilon)$, $y_2(0) = 0$, ${\bf y}_3 = {\bf O}_{n-2}(\varepsilon)$
and with the period $T = \frac{2 \pi}{\omega_0} + O(\varepsilon)$.
\end{theorem}

\section{Hidden attractors localization in Chua circuit via the describing function method}

In this section we apply the above approach for hidden attractors localization in Chua circuit.
Let us write Chua system \eqref{chuasys} in the Lur'e form
\eqref{sys_gen}  (see, e.g. \citep{LeonovKV-2011-PLA})
with
\begin{equation} \label{ChuaLourie}
\begin{aligned}
	{\bf P} &= \left(
    \begin{array}{ccc}
       -\alpha(m_1+1) & \alpha & 0 \\
       1 & -1 & 1 \\
       0 & -\beta & -\gamma \\
    \end{array}
   \right), \, {\bf q} = \left( \begin{array}{c} -\alpha \\ 0 \\ 0 \\ \end{array}\right), &\\
   {\bf r} &= \left( \begin{array}{c} 1 \\ 0\\ 0 \\ \end{array} \right), \quad
   \psi(\sigma) = (m_0 - m_1) \, {\rm sat}(\sigma).
\end{aligned}
\end{equation}

Introduce a coefficient $k$ and a small parameter $\varepsilon$, and represent
\eqref{ChuaLourie} in the form \eqref{sys_gen_phi_0} with
\begin{equation} \label{ChuaLphi}
\begin{aligned}
	&{\bf P_0} ={\bf P}+k{\bf qr}^*= \left(
     \begin{array}{ccc}
       -\alpha(m_1+1+k)& \alpha & 0 \\
       1 & -1 & 1 \\
       0 & -\beta & -\gamma \\
     \end{array}
   \right), \\
   & \varphi(\sigma) = \psi(\sigma) - k\sigma = (m_0-m_1) \, {\rm sat}(\sigma) - k\sigma,
\end{aligned}
\end{equation}
and $\lambda^{{\bf P}_0}_{1,2} = \pm i\omega_0$, $\lambda^{{\bf P}_0}_{3} = -d < 0$.

By the non-singular linear transformation ${\bf x} = \bf{Sy}$
system \eqref{ChuaLphi} is reduced to the form \eqref{sys_hlin}
\begin{equation} \label{ChuaDiag}
  \frac{d{\bf y}}{dt} = {\bf A}{\bf y} + {\bf b}\varepsilon\varphi({\bf u}^*{\bf y}),
\end{equation}
where
\[
   {\bf A} = \left(
    \begin{array}{ccc}
       0 & -\omega_0 & 0 \\
       \omega_0 & 0 & 0 \\
       0 & 0 & -d \\
     \end{array}
   \right),
   \
   {\bf b} = \left( \begin{array}{c} b_1 \\ b_2 \\ 1 \\ \end{array}\right),
   \
   {\bf c} = \left( \begin{array}{c} 1 \\ 0\\ -h \\ \end{array} \right).
\]
The transfer function $W_{{\bf A}}(p)$ of system \eqref{ChuaDiag} can be represented as
\[
  W_{{\bf A}}(p) = \dfrac{-b_1p+b_2\omega_0}{p^2+\omega_0^2}+\dfrac{h}{p+d}.
\]
Further, using the equality of transfer functions of systems \eqref{ChuaLphi} and \eqref{ChuaDiag}
one can obtain
\[
  W_{{\bf A}}(p)={\bf r}^*({\bf P}_0-p{\bf I})^{-1}{\bf q}.
\]
This implies the following relations
\begin{equation}\label{chua_kdh}
  \begin{aligned}
  & k = \frac{-\alpha(m_1+m_1\gamma+\gamma)+\omega_0^2-\gamma-\beta}{\alpha(1+\gamma)}, \\ 
  & d = \frac{\alpha+\omega_0^2-\beta+1+\gamma+\gamma^2}{1+\gamma}, \\ 
  & h = \frac{\alpha(\gamma+\beta-(1+\gamma)d+d^2)}{\omega_0^2+d^2}, \\ 
  & b_1 = \frac{\alpha(\gamma+\beta-\omega_0^2-(1+\gamma)d)}{\omega_0^2+d^2}, \\ 
  & b_2 = \frac{\alpha\big((1+\gamma-d)\omega_0^2+(\gamma+\beta)d\big)}{\omega_0(\omega_0^2+d^2)}. 
  \end{aligned}
\end{equation}

Since by the non-singular linear transformation ${\bf x} = \bf{Sy}$
system \eqref{ChuaLphi} can be reduced to the form \eqref{ChuaDiag}
for the matrix ${\bf S}$ the following relations
\begin{equation}\label{chua_defS}
  {\bf A} = {\bf S}^{-1}{\bf P_0S}, \quad {\bf b}={\bf S}^{-1}{\bf q}, \quad {\bf c}^*={\bf r}^*{\bf S}.
\end{equation}
are valid.
After solving these matrix equations, one can obtain the transformation matrix
\[
  {\bf S} = \left(
    \begin{array}{ccc}
      s_{11} & s_{12} & s_{13} \\
      s_{21} & s_{22} & s_{23} \\
      s_{31} & s_{32} & s_{33} \\
    \end{array}
  \right),
\]
where
\begin{align*}
	& s_{11}=1, \qquad  s_{12}=0, \qquad s_{13}=-h, \\
	& s_{21}=m_1+1+k, \qquad s_{22}=-\frac{\omega_0}{\alpha}, \\
	& s_{23}=-\frac{h(\alpha(m_1+1+k)-d)}{\alpha}, \\
	& s_{31}=\frac{\alpha(m_1+k)-\omega_0^2}{\alpha}, \\
	& s_{32} = -\frac{\alpha(\beta+\gamma)(m_1+k)+\alpha\beta-\gamma\omega_0^2}{\alpha\omega_0}, \\
	& s_{33} = h\frac{\alpha(m_1+k)(d-1)+d(1+\alpha-d)}{\alpha}.
\end{align*}

Using Theorem \ref{th_stable}
one obtains the initial data
\begin{equation} \label{initialdata}
  {\bf x}(0) = {\bf S}{\bf y}(0) =
  {\bf S} \, \left( \begin{array}{c} a_0 \\ 0 \\ 0 \\ \end{array} \right) =
  \left( \begin{array}{c} a_0 \, s_{11} \\ a_0 \, s_{21} \\ a_0 \, s_{31} \\ \end{array} \right).
\end{equation}

Back to Chua system denotations, for the determination of the initial data of starting
solution for multistage procedure, it can be obtained
\begin{equation} \label{initialdataChua}
\begin{aligned}
  & x(0) = a_0, \quad y(0) = a_0(m_1 + 1 + k), \, \\
  & z(0)=a_0 \, \frac{\alpha(m_1 + k) - \omega_0^2}{\alpha}.
\end{aligned}
\end{equation}

Consider system \eqref{chuasys} with the parameters
\begin{equation}\label{chuasys:params1}
\begin{aligned}
	&\alpha = 8.4562,\quad \beta = 12.0732,\quad \gamma = 0.0052, \\
	&\quad m_0 = -0.1768,\quad m_1 = -1.1468.
\end{aligned}
\end{equation}
Note that for the considered values of parameters there
are three equilibria in the system:
the zero equilibrium $F_0 = (0, \, 0, \, 0)$ is a stable focus-node
and two symmetric equilibria
\[
	S_{\pm} = \pm \left(
\frac{m_1 - m_0}{m_1 + \frac{\beta}{\beta+\gamma}}, \,
\frac{\gamma (m_1 - m_0)}{(\gamma + \beta) m_1 + \beta}, \,
-\frac{\beta (m_1 - m_0)} {(\gamma + \beta) m_1 + \beta}\right)
\]
are saddle-foci with one-dimensional unstable manifolds.

\begin{figure}[!ht]
  \centering
  	\includegraphics[width=0.45\textwidth]{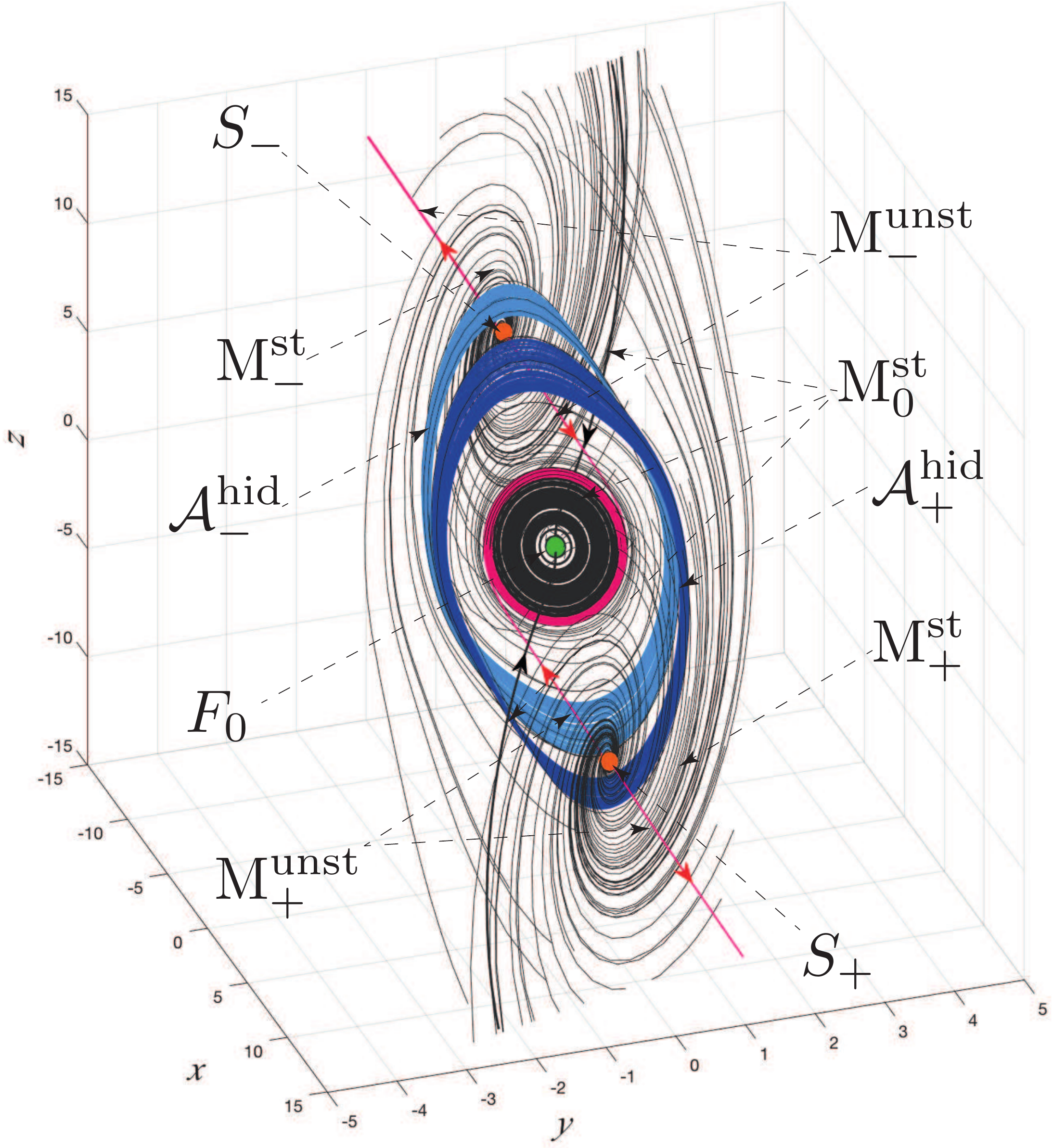}
  \caption{Two symmetric hidden chaotic attractors
  ($\mathcal{A}_{\pm}^{\rm hid}$ - blue domains)
  in the classical Chua system \eqref{chuasys}:
  trajectories (red) from unstable manifolds ${\rm M_{\pm}^{\rm unst}}$
  of two saddle points $S_{\pm}$ are either attracted to
  locally stable zero equilibrium $F_0$, or tend to infinity;
  trajectories (black) from stable manifolds ${\rm M_{0,\pm}^{\rm st}}$ tend to
  $F_0$ or $S_{\pm}$;
  $\alpha = 8.4562$, $\beta = 12.0732$, $\gamma = 0.0052$, $m_0 = -0.1768$, $m_1 = -1.1468$.
  }
  \label{fig:hidden01}
\end{figure}

Let us try to apply the DFM and define an initial data for periodic oscillation.
Using \eqref{eq:omega0} and \eqref{eq:harm-lin-coef} for parameters \eqref{chuasys:params1}
one obtains following starting frequency and a coefficient of harmonic:
\begin{equation}\label{param:omega0k:vals1}
	\omega_0 = 2.0392, \quad k = 0.2098.
\end{equation}
Assuming $a \geq 1$, describing function \eqref{eq:descr-func} and its derivative
for Chua system \eqref{chuasys} can be rewritten as follows:
\begin{align*}
	&\Phi(a) = 2 (m_0 - m_1) \left[\frac{\pi a}{2} + \sqrt{1 - \frac{1}{a^2}}
	- a \arccos\frac{1}{a}\right] - \pi a k,\\
  & \frac{d\Phi(a)}{da} = 2 (m_0 - m_1) \left[\frac{\pi}{2} - \frac{1}{a}\sqrt{1 - \frac{1}{a^2}}
  - \arccos\frac{1}{a}\right] - \pi k.
\end{align*}
For parameters \eqref{chuasys:params1} and \eqref{param:omega0k:vals1}
one obtains initial amplitude $a_0 = 5.8576$ that satisfies the conditions
of Theorem~\ref{th_stable}. Thus, by \eqref{initialdataChua} initial data
for the oscillation are as follows
\begin{equation}\label{initdata:hidden:val1}
	x(0) = 5.8576, \quad y(0) = 0.3694, \quad z(0) = -8.3686.
\end{equation}
In our numerical experiments we skip the multistep procedure based on
the small parameter method and apply initial data
\eqref{initdata:hidden:val1} for hidden attractors
localization in the initial system (i.e., system \eqref{chuasys}
in the form \eqref{sys_gen_p0}, $\varepsilon = 1$).
It turns out that in this case this is enough for localization of
two symmetric hidden chaotic attractors $\mathcal{A}_{\pm}^{\rm hid}$
in the Chua system (see Fig.~\ref{fig:hidden01}).
For attractor $\mathcal{A}_{-}^{\rm hid}$ one should take
symmetric initial data $x(0) = -5.8576$ $y(0) = -0.3694$, $z(0) = 8.3686$.

\begin{figure}[!ht]
  \centering
  	\includegraphics[width=0.49\textwidth]{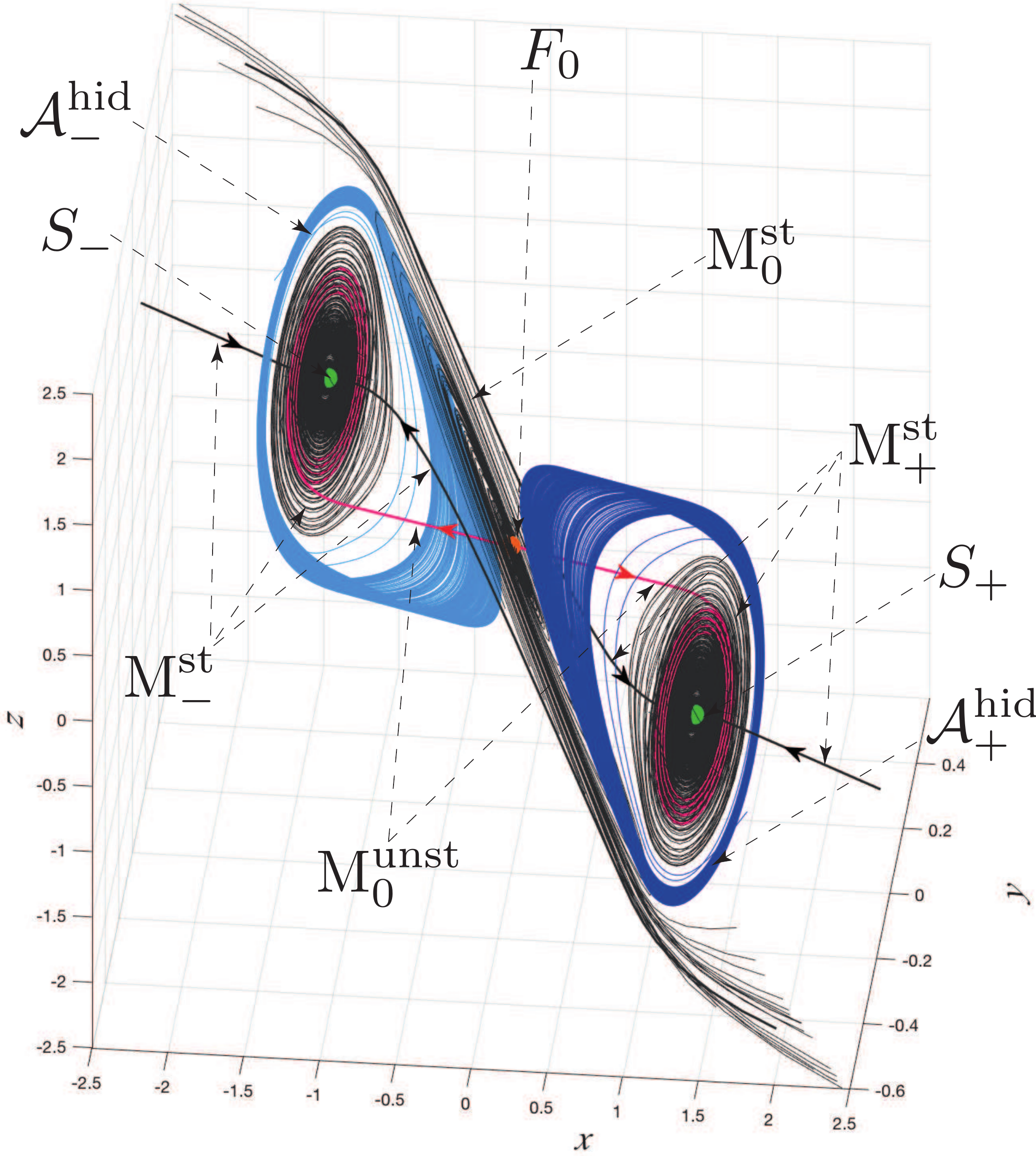}
  \caption{Two symmetric hidden chaotic attractors
  ($\mathcal{A}_{\pm}^{\rm hid}$ - blue domains)
  in the classical Chua system \eqref{chuasys}:
  trajectories (red) from unstable manifold ${\rm M_{0}^{\rm unst}}$
  of the saddle points $F_0$ are attracted to locally stable equilibria $S_{\pm}$;
  trajectories (black) from stable manifolds ${\rm M_{0,\pm}^{\rm st}}$ are attracted to
  $F_0$ or $S_{\pm}$; $\alpha = 8.4$, $\beta = 12$, $\gamma = -0.005$, $m_0 = -1.2$, $ m_1 = -0.05$.
  }
  \label{fig:hidden02}
\end{figure}

Consider system \eqref{chuasys} with another values of the parameters
\begin{equation}\label{chuasys:params2}
\begin{aligned}
	&\alpha = 8.4,\quad \beta = 12,\quad \gamma = -0.005, \\
	&\quad m_0 = -1.2,\quad m_1 = -0.05.
\end{aligned}
\end{equation}
Note that for the considered values of parameters
the zero equilibrium $F_0$ is a saddle-focus with one-dimensional unstable manifold
and two symmetric equilibria $S_{\pm}$
are stable focus-nodes.
Again let us apply the DFM and define an initial data for periodic oscillation.
Note that equation \eqref{eq:omega0} for parameters \eqref{chuasys:params2}
has two positive solutions and by \eqref{eq:harm-lin-coef} we
obtain following starting frequencies and coefficients of harmonic:
\begin{equation}\label{param:omega0k:vals2}
	\omega_0 = 2.0260, \quad k = -0.8890
\end{equation}
and
\begin{equation}\label{param:omega0k:vals3}
	\omega_0 = 3.2396, \quad k = -0.1244.
\end{equation}
For parameters \eqref{chuasys:params2} and \eqref{param:omega0k:vals2}
one obtains initial amplitude $a_0 = 1.5187$ that satisfies the conditions
of Theorem~\ref{th_stable}. Thus, by \eqref{initialdataChua} initial data
for the oscillation are as follows
\begin{equation}\label{initialdata:hidden:val2}
	x(0) = 1.5187, \quad y(0) = 0.0926, \quad z(0) = -2.1682.
\end{equation}
Using these initial data for original system \eqref{chuasys}
it is possible to localize two symmetric hidden chaotic attractors
$\mathcal{A}_{\pm}^{\rm hid}$ (see Fig~\ref{fig:hidden02}).

\begin{figure}[!ht]
  	\centering
    \includegraphics[width=0.5\textwidth]{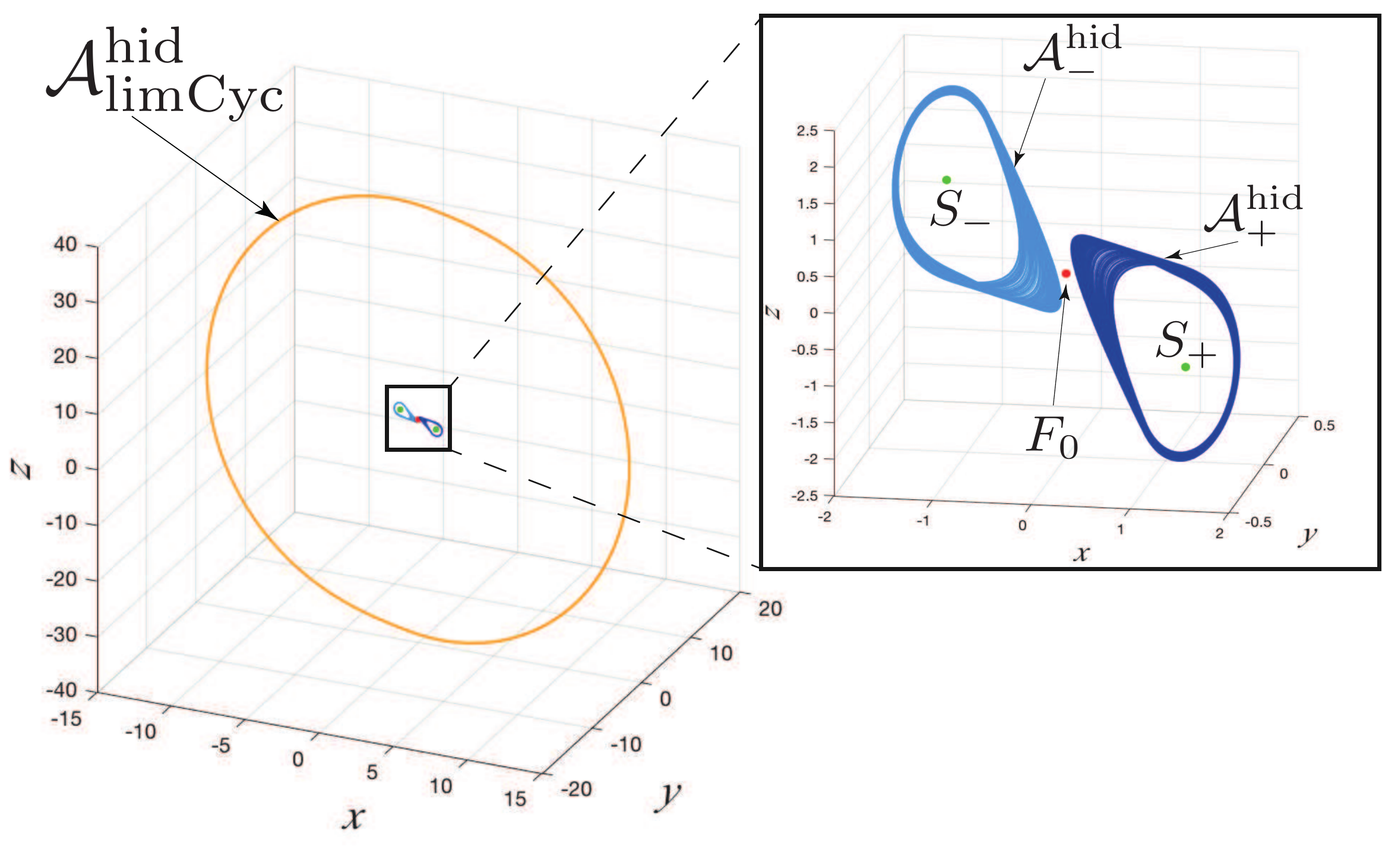}
    \caption{Co-existence of  hidden periodic attractor:
    the stable limit cycle $\mathcal{A}^{\rm hid}_{\rm limCyc}$,
    and two symmetric hidden chaotic attractors $\mathcal{A}^{\rm hid}_{\pm}$
    in Chua system \eqref{chuasys} with parameters
  	$\alpha = 8.4$, $\beta = 12$,$\gamma = -0.005$,
  	$m_0 = -1.2$, $ m_1 = -0.05$.}
    \label{fig:hidden}
\end{figure}

For parameters \eqref{chuasys:params2} and \eqref{param:omega0k:vals3}
one obtains initial amplitude $a_0 = 11.7546$ (also satisfies the conditions
of Theorem~\ref{th_stable}) which by \eqref{initialdataChua} yields
the following initial data
\begin{equation}\label{initialdata:hidden:val2}
	x(0) = 11.7546, \,\, y(0) = 9.7044, \,\, z(0) = -16.7367.
\end{equation}
These initial data for original Chua system \eqref{chuasys} allows to localize
a hidden periodic attractor: stable limit cycle $\mathcal{A}^{\rm hid}_{\rm limCyc}$ (see Fig. \ref{fig:hidden}).
Thus, in this configuration despite the trivial attractors, i.e. equilibria $S_{\pm}$,
for system \eqref{chuasys} with parameters \eqref{chuasys:params2}
we obtain the co-existence of hidden periodic attractor (stable limit cycle)
and two symmetric hidden chaotic attractors.

\section*{Conclusions}
In this paper we discuss the use of describing function method for
searching periodic oscillations in its application to the famous Chua circuit.
Despite the fact that DFM is an approximate analytical method
(which does not guarantee the true results),
the application of DFM to the Chua system allows us
to localize hidden chaotic and periodic attractors.
In particular, for certain values of parameters we obtain
a new configuration of co-existing hidden attractors
(two symmetric chaotic and stable limit cycle) in the Chua system.

\section{Acknowledgments}
This work was supported by the Russian Science Foundation
(14-21-00041).


\end{document}